%% file: paper.tex
\documentclass[12pt]{article}
\usepackage[dvips]{graphics,epsfig}

\title{$SU(2)/Z_2$ symmetry of the BKT transition and 
	twisted boundary condition}

\author{Kiyohide Nomura$^{1}$	and Atsuhiro Kitazawa$^{1,2}$	\\
        {\em $^1$ Department of Physics, Kyushu University,} \\
        {\em Fukuoka 812-81, Japan.} \\
        {\em $^2$ Department of Physics, Tokyo Institute of Technology,} \\
        {\em Tokyo 152, Japan.} \\
}

\date{\today}

\begin{document}

\maketitle

\begin{abstract}
Berezinskii-Kosterlitz-Thouless (BKT) transition, 
the transition of the 2D sine-Gordon model, plays an important role 
in the low dimensional physics.  We relate the operator content of the BKT 
transition to that of the SU(2) Wess-Zumino-Witten model, using twisted 
boundary conditions.  With this method, in order to determine the BKT
critical point, we can use the level crossing
of the lower excitations than the periodic boundary case, 
thus the convergence to the transition point is highly improved.  
Then we verify the efficiency of this method by applying to the S=1,2 
spin chains.  
\end{abstract}

PACS numbers: 75.40.Cx, 05.70.Fh, 11.10.Hi, 75.10.Jm

\pagebreak

\section{Introduction}

Berezinskii-Kosterlitz-Thouless (BKT) transition 
\cite{Berezinskii, Kosterlitz-T,Kosterlitz} plays an important role in
2D classical and one-dimensional (1D) quantum systems, such as the 2D
XY spin model, 2D helium film, 2D superconducting film, 
roughening transition, 1D quantum spin
models and 1D electron systems.  
Furthermore, BKT transition is a typical instability of the
Tomonaga-Luttinger liquid, such as the Mott transition in 1D metal 
\cite{Giamarchi}.  

It was a difficult problem to determine numerically the critical point and
the universality class of the BKT transition, because of logarithmic
corrections and the slow divergence of the inverse correlation length
(or the energy gap) \cite{Gupta}.  
Moreover, it is hard to distinguish the BKT transition from the second
order transition, by using the conventional finite-size-scaling 
method \cite{Seiler,Edwards}.  
In fact, for the exactly solvable S=1/2 XXZ spin chain, where the
transition at $\Delta=1$ is of the BKT type, the finite-size scaling
method leads to false conclusions \cite{Solyom}.  
This problem was successfully resolved by the level
spectroscopy \cite{Nomura}, based on the renormalization group calculation
and the $SU(2)/Z_2$ symmetry in the BKT transition.  

The SU(2) symmetry inherent in the BKT transition was pointed out by
Halpern \cite{Halpern} and 
Banks {\em et al.}\cite{Banks}.  They showed the equivalence
between the SU(2) massless Thirring model and the theory of the bosons
consisting of a free field plus a $\beta^2=8\pi$ sine-Gordon model,
which corresponds to the BKT line.  
Another approach was proposed by Ginsparg \cite{Ginsparg}, based on
the $c=1$ conformal field theory: modding out the SU(2) symmetry model
by $Z_2$ symmetry gives the structure of the BKT multicritical point. 

In this paper we directly relate the $SU(2)/Z_2$ structure of the
operator content of the BKT transition 
to that of the $k=1$ SU(2) Wess-Zumino-Witten (WZW) model, 
using twisted boundary conditions.  
With this method, in order to determine the BKT
critical point, we can use the level crossing
of the lower excitations than the periodic boundary case, 
thus the convergence to the transition point is highly improved.  
We then apply the method to the S=1,2 spin chains.

\section{BKT transition and twisted boundary condition}

\subsection{On the Gaussian line}

We consider the 2D Gaussian model defined as the Lagrangian
\begin{equation}
	{\cal L}_0= \frac{1}{2 \pi K} (\nabla \phi)^2.
	\label{eq:Gauss-Lagrangian}
\end{equation}
We compactify $\phi$ on a circle as $\phi \equiv \phi + 2\pi/\sqrt{2}$.  
We introduce the dual field $\theta$ to $\phi$ defined as 
\begin{equation}
	\partial_x \phi = - \partial_y (iK \theta), \;
	\partial_y \phi =  \partial_x (iK \theta),
\end{equation}
which has the periodic nature $\theta \equiv \theta+ 2\pi/\sqrt{2}$.  

The vertex operator defined as 
\begin{equation}
	O_{m,n} = :\exp(i m \sqrt{2} \phi) \exp(i n \sqrt{2} \theta):,
\end{equation}
has a scaling dimension $x_{m,n}$ and a conformal spin $s_{m,n}$
\begin{equation}
	x_{m,n}(K)= \frac{1}{2} \left( m^2 K + \frac{n^2}{K} \right), \; 
	s_{m,n}= mn.
\end{equation}
There are the other type fields, i.e., the current fields 
\begin{equation}
	\frac{2 i}{\sqrt{K}} \partial \phi,	\;
	\frac{2 i}{\sqrt{K}} \bar{\partial} \phi,
\end{equation}
which have $x=1,s=\pm1$, and the marginal field 
\begin{equation}
	{\cal M} 
	= -\frac{4}{K} \partial \phi \bar{\partial} \phi
	= -\frac{1}{K} \left((\partial_x \phi)^2+(\partial_y \phi)^2 \right)
\end{equation}
which has $x=2,s=0$.  There exist also descendant fields of them.  

About the symmetry, the Gaussian model (\ref{eq:Gauss-Lagrangian}) 
is invariant under $\phi \rightarrow \phi +const.,
\theta \rightarrow \theta +const.$, which means 
the $U(1) \times U(1)$ continuous symmetry.  
There are also the discrete $Z_2$ symmetries, 
$T: (z, \phi, \theta) \rightarrow (z, -\phi, -\theta)$ and
$C: (z, \phi, \theta) \rightarrow (\bar{z}, \phi, -\theta)$  
(we can also define the redundant symmetry
$P=CT: (z, \phi, \theta) \rightarrow (\bar{z}, -\phi, \theta)$).
The Gaussian model is invariant under the dual transformation 
$K \leftrightarrow 1/K$, $\phi \leftrightarrow \theta$.  
The self dual point $K=1$ is nothing but the $k=1$ $SU(2) \times SU(2)$ 
Wess-Zumino-Witten model, whose symmetry structure is apparent in the 
conformal dimensions of the vertex, the current, the marginal operators 
and their descendants.  
At the point $K=1$, the number of the marginal operator ($x=2,s=0$) is 9.  

The $K=4$ is the BKT transition point, where 
the number of the marginal operator ($x=2,s=0$) is 5.  
The correspondence of the scaling dimensions for the WZW point ($K=1$) 
with those for the BKT point ($K=4$) is given by
\begin{equation}
	x_{2m,n}(1) = x_{m,2n} (4),
	\label{eq:operator-corr-periodic}
\end{equation}
and
\begin{equation}
	x_{m,2n}(1) = x_{n,2m} (4).
	\label{eq:operator-corr-periodic-dual}
\end{equation}
In short, there is a partial correspondence of the operator content 
of these two models, reflecting the $Z_2$ symmetry difference.  

Fortunately, introducing a half-integer magnetic charge, one can obtain 
the full correspondence of the $K=1$ and $K=4$ Gaussian models.  
With twisted boundary conditions on a cylinder, or the cut from $0$ 
to infinity on the complex plane, one can set the arbitrary charges 
at $\pm \infty$ (on the cylinder) or at $0, \infty$ (on the complex plane) 
\cite{Blote-Cardy-Nightingale,Destri,Kitazawa} (see Appendix A).  
The relation between the twisted boundary condition and 
the change of the charges has been observed 
in the Bethe-Ansatz solvable S=1/2 XXZ spin chain 
\cite{Alcaraz-Barber-Batchelor}.  
Therefore, with the twist angle $\Phi=\pi$, one can introduce 
a half-magnetic charge.  In that case, we can set
\begin{equation}
	x_{m,n}^{TBC}(K) = x_{m+1/2,n} (K).
\end{equation}
Then, we obtain
\begin{equation}
	x_{2m+1,n}(1) = x_{m,2n}^{TBC} (4),
	\label{eq:operator-corr-twist}
\end{equation}
and
\begin{equation}
	x_{m,2n}^{TBC}(1) = x_{n,2m+1} (4),	\;
	x_{m,2n+1}^{TBC}(1) = x_{n,2m+1}^{TBC}(4),
	\label{eq:operator-corr-twist-dual}
\end{equation}
so that we can see the correspondence of all the operator content  
of $K=1$ and $K=4$ points, considering twisted boundary conditions.

\subsection{Renormalization}

Next we introduce the sine-Gordon type interaction, that is,
\begin{equation}
	{\cal L}_I = \frac{y_\phi}{2 \pi \alpha^2} :\cos \sqrt{8} \phi:	\;
	 \mbox{(for $K=1$)},	\;
	\frac{y_\phi}{2 \pi \alpha^2} :\cos \sqrt{2} \phi:	\;
	\mbox{ (for $K=4$), }
\end{equation}
(where $\alpha$ is a short-distance (ultraviolet) cut-off).  
Here we assume that there is no 
$\cos \sqrt{2} \phi$ interaction term for the $K=1$ case, 
by the symmetry reason 
(discrete symmetry $\phi \rightarrow \phi +\pi/\sqrt{2}$ 
which corresponds to ``translation by one site'' in the half-odd integer 
spin chain \cite{Affleck}) 
or adjusting parameters.  
With the simple transformation 
$\phi \rightarrow 2\phi, \theta \rightarrow \theta/2$ 
or $m \rightarrow 2m, n \rightarrow n/2$, 
the sine-Gordon model for $K=4$ becomes equivalent to 
the sine-Gordon model for $K=1$.  

About the symmetry, the U(1) symmetry of $\phi$ is explicitly broken
to the discrete symmetry 
$\phi \rightarrow \phi + 2\pi/\sqrt{8} (K=1)$ or 
$\phi \rightarrow \phi + 2\pi/\sqrt{2} (K=4)$.  
Furthermore, the Lagrangian (\ref{eq:Gauss-Lagrangian}) is invariant under 
$\phi \rightarrow \phi + \pi/\sqrt{8}, y_\phi \rightarrow -y_\phi (K=1)$ or 
$\phi \rightarrow \phi + \pi/\sqrt{2}, y_\phi \rightarrow -y_\phi (K=4)$.  

Under a change of cut-off $\alpha \rightarrow e^l \alpha$, 
the renormalization group equations for the sine-Gordon model are 
\cite{Kosterlitz}
\begin{eqnarray}
	\frac{d y_0(l)}{dl} &=& - y_\phi^2 (l)	\nonumber	\\
	\frac{d y_\phi(l)}{dl} &=& - y_\phi (l) y_0 (l)
\end{eqnarray}
where $K=1+y_0/2$(around $K=1$) or $K=4(1+y_0/2)$ ($K=4$).  
For the finite system, $l$ is related to the system size $L$ 
by $l= \log (L/\alpha)$.  
There are three critical lines; $y_\phi=0$ corresponding to the
Gaussian fixed line, and $y_\phi= \pm y_0 (y_0>0)$ corresponding to
the BKT lines.  
On the BKT lines, the couplings behave as 
$y_0(l) = \pm y_\phi(l) = 1/(l + 1/y_0(0)) = 1/\log (L/L_0)$.  
In the region between the two BKT lines, all the
points will be renormalized to the Gaussian fixed line, so they are
massless.  The other region is massive, except on the Gaussian fixed
line.  

Although the renormalization flow is the same between the $K=1$ and
the $K=4$ cases, the operator content is different, and in general the
correspondences (\ref{eq:operator-corr-periodic-dual}), 
(\ref{eq:operator-corr-twist-dual}) are not satisfied, 
since the duality relation $K \leftrightarrow 1/K$ does not hold 
on the BKT lines $y_\phi= \pm y_0$.  
 Fortunately, the correspondences (\ref{eq:operator-corr-periodic}),
(\ref{eq:operator-corr-twist}) remain correct after the renormalization, since
the operator product expansion (OPE) structure of both cases is
the same, for example,
\begin{eqnarray}
	K=1	& &	K=4	\nonumber	\\
	\langle \cos \sqrt{2} \phi \cos \sqrt{8} \phi \cos \sqrt{2} \phi \rangle
	\;	&\leftrightarrow&
	\langle \cos \phi /\sqrt{2} \cos \sqrt{2} \phi \cos \phi / \sqrt{2}\rangle
	\nonumber	\\
	\langle \sin \sqrt{2} \phi \cos \sqrt{8} \phi \sin \sqrt{2} \phi \rangle
	\;	&\leftrightarrow&
	\langle \sin \phi /\sqrt{2} \cos \sqrt{2} \phi \sin \phi / \sqrt{2}\rangle
	\nonumber	\\
	\langle \cos \sqrt{8} \phi {\cal M} \cos \sqrt{8} \phi \rangle
	\;	&\leftrightarrow&
	\langle \cos \sqrt{2} \phi {\cal M} \cos \sqrt{2} \phi \rangle
\end{eqnarray}
which can be seen by the mapping 
$\phi \rightarrow 2\phi, \theta \rightarrow \theta/2$ 
(similarly, higher OPE's are the same).  
The renormalized scaling dimensions are determined by these OPE 
\cite{Giamarchi-Schulz,Nomura} (see Appendix B).  
Therefore, using twisted boundary conditions, 
we can explicitly see the SU(2) structure inherent in the BKT
transition.  

In Table 1 we summarize the operator content of the WZW ($K=1$) model
and that of the BKT ($K=4$) model, with renormalized scaling 
dimensions.  
Note that at the BKT critical line ($y_\phi(l) =y_0(l)$), 
there are degeneracies of the excitations corresponding 
to $x^{p1,p2},x^{p3}$ or $x^{m0},x^{m3},x^{m7,m8}$ or $x^{m2},x^{m5,m6}$,
respectively, reflecting the SU(2) ($SU(2)/Z_2$) symmetry 
(for the $y_\phi (l) = - y_0(l)$ branch, 
the role of operators $x^{p0} \leftrightarrow x^{p3}, \;
x^{m5,m6} \leftrightarrow x^{m7,m8}$ interchanges).  
Thus the level crossing of them can be used 
to determine the BKT critical point.  

In practical systems, there are corrections from the descendant fields of
the identity operator {\bf 1}. The most important irrelevant fields
in them are  
$L_{-2}\bar{L}_{-2} {\bf 1}, (L_{-2}^2+\bar{L}_{-2}^2) {\bf 1}$ with scaling 
dimension $x=4$ \cite{Cardy86,Reinicke}.  
With the twisted boundary conditions, we can use the level crossing of
the lower excitations than that with only the periodic boundary
conditions.  
Since the amplitude of the corrections from the irrelevant field 
becomes smaller with the lower excitations \cite{Reinicke}, 
thus the convergence to the transition point is highly improved.

\section{Physical examples}

\subsection{ S=1 XXZ chain}

As a physical example, let us consider the S=1 XXZ spin chain,
described by the Hamiltonian:
\begin{equation}
	H= \sum_{j=1}^L h_{j,j+1}, \;
	h_{j,j+1}=  \frac{1}{2}(S^+_j S^-_{j+1}+ S^-_j S^+_{j+1}) 
	+ \Delta S^z_j S^z_{j+1},
	\label{eq:S=1-Hamiltonian}
\end{equation}
where we assume $L$ is even and periodic boundary conditions.  
This model is invariant under spin rotation around the $z$-axis, 
translation($T_R:S^{x,y,z}_j \rightarrow S^{x,y,z}_{j+1}$), 
space inversion ($P:S^{x,y,z}_j \rightarrow S^{x,y,z}_{L-j+1}$),
spin reversal ($T:S^z_j \rightarrow -S^z_j, S^{\pm}_j \rightarrow S^{\mp}_j$).  
Therefore, energy eigenstates are characterized by the $z$-component of the
total spin ($S^z_T = \sum S^z_j$), wavenumber ($q= 2\pi k/L$), 
parity ($P = \pm 1$), spin reversal ($T=\pm 1$).  
Note that with the unitary transformation $\exp (\pi i \sum j S^z_j)$, 
spin operators change as 
$S^z_j \rightarrow S^z_j, S^{\pm}_j \rightarrow (-1)^j S^{\pm}_j$, 
thus the momentum of the excitation in the odd $S^z_T$ changes 
$q \rightarrow q+\pi$, 
and the sign of the XY term in (\ref{eq:S=1-Hamiltonian}) reverses.  

About the spin $S$ XXZ chain, Haldane \cite{Haldane} 
discussed that for the integer $S$
case, there are two transition points $\Delta_{c1}<1<\Delta_{c2}$,
where the XY-Haldane transition $\Delta_{c1}$ is of the BKT type, and
the Haldane-N\'eel transition point $\Delta_{c2}$ is of the 2D Ising
universality class,  
in contrast to the half-odd integer $S$ case where only one
transition point exists at $\Delta=1$ of the SU(2) WZW type.  

Numerically Botet and Jullien estimated 
$\Delta_{c1} \approx 0.1$ \cite{Botet}, 
Sakai and Takahashi $\Delta_{c1} = -0.01 \pm 0.03$ \cite{Sakai}, 
and Yajima and Takahashi $\Delta = 0.069 \pm 0.003$ \cite{Yajima}.  
Recently, the authors found  $\Delta_{c1}=0$ and checked the
universality class \cite{Kitazawa-Nomura-Okamoto} using 
the level spectroscopy \cite{Nomura}.  
We found the exact degeneracy (at least within numerical accuracy) 
between the $S^z_T = \pm 4,q=0, P=1$ and the $S^z_T =0, q=0, P=T=1$ 
excitations.

These results are obtained under periodic boundary conditions.  
To consider a twisted boundary condition, in (\ref{eq:S=1-Hamiltonian})
we replace the boundary term between the L-th site and the 1st site, 
\begin{equation}
	h_{L,1}=  \frac{1}{2}(S^+_L S^-_1 \exp (-i\Phi) 
	+ S^-_L S^+_1 \exp(i\Phi)) + \Delta S^z_L S^z_1.
	\label{eq:TBC1}
\end{equation}
However, with this boundary condition the Hamiltonian
(\ref{eq:S=1-Hamiltonian}) is not translational invariant.  
The translational invariance can be restored, still maintaining the
total angle $\Phi$, by twisting all neighboring bonds in the chain by
an angle $\Phi/L$ with the next unitary transformation 
\begin{eqnarray}
	U_{\Phi} &=& \exp \left( i \frac{\Phi}{L} \sum_{j=1}^{L} 
	(j-\frac{1}{2}) S^z_j \right),	\nonumber	\\
	U_\Phi S^{\pm}_j U_\Phi^{-1} &=& S^{\pm}_j \exp (\pm i (j-1/2)\Phi/L),	\;
	U_\Phi S^{z}_j U_\Phi^{-1} = S^{z}_j,
	\label{eq:TBC2}
\end{eqnarray}
(in $U_{\Phi}$ we use $(j-1/2)S^z_j$ for the compatibility of 
the definition of $P$). Thus, we obtain 
\begin{equation}
	H_t(\Phi)= \sum h^{t}_{j,j+1}, 	\;
	h^{t}_{j,j+1} = \frac{1}{2}(S^+_j S^-_{j+1} e^{-i\Phi/L} 
	+ S^-_j S^+_{j+1} e^{i\Phi/L}) 
	+ \Delta S^z_j S^z_{j+1}.
	\label{eq:S=1-Twisted-Hamiltonian}
\end{equation}
Under this boundary condition, we had better modify the definition of the
translational operator $T_R^t \equiv \exp(i \Phi S^z_T/L) T_R$ 
(see Appendix C).  
Although there are no discrete symmetries like $P,T$ in general $\Phi$, 
for the special angle $\Phi=\pi$, the system
(\ref{eq:S=1-Twisted-Hamiltonian}) is invariant under the
discrete symmetries $U_{2\pi} P,U_{2\pi} T$.  
Moreover, it was shown that $U_{2\pi} P,U_{2\pi} T$ are the good
quantum numbers \cite{Kitazawa-Nomura} 
characterizing the generalized $Z_2 \times Z_2$
symmetries \cite{Oshikawa} (see also Appendix C).  
Thus the twist angle $\Phi= \pi$ has the special meaning 
in addition to the one discussed in section 2.  
Hereafter we call the twisted boundary condition with
$\Phi=\pi$ as TBC and the periodic boundary conditions as PBC.  

The energy eigenvalues $E_n$ are related to the scaling dimension $x_n$ as
\begin{equation}
	E_n (L)- E_g (L) = \frac{2 \pi v x_n}{L},
\end{equation}
where $v$ is the spin wave velocity \cite{Cardy84}.
And the conformal anomaly $c$ is related to the ground state energy 
\cite{Blote-Cardy-Nightingale, Affleck86}
\begin{equation}
	E_g (L) = e_g L - \frac{\pi v c}{ 6 L}.
\end{equation}

In Fig.1 we show the excitations corresponding to 
the scaling dimension $x=1/2$.  
The excitation $S^z_T=0, q=0, U_{2\pi}P=U_{2\pi}T=-1$ under TBC 
is exactly degenerate with excitations $S^z_T=\pm2,q=0,P=1$ under PBC 
at $\Delta=0$.  
Fig.2 shows the excitations corresponding to the scaling dimension $x=2$.  
At $\Delta=0$, the excitations $S^z_T= \pm2, q=0, U_{2\pi}P=-1 $ under TBC  
are exactly degenerate with excitations $S^z_T=0,q=0,P=T=1$ and 
$S^z_T= \pm4,q=0,P=1$ under PBC, 
whereas the excitations $S^z_T= \pm2, q=0, U_{2\pi}P=1 $ under TBC  
are exactly degenerate with the excitation $S^z_T=0,q=0,P=T=-1$ under PBC.  
Fig.3 shows the excitations with scaling dimension $x=1$, 
$S^z_T=0,q= \pm 2\pi/L$ with PBC, $S^z_T=\pm 2,q= \pm 2\pi/L$ with TBC.  
They are exactly degenerate at $\Delta=0$.  

Next we discuss the universality relations.  
The conformal anomaly is estimated $c=1.000$ in 
\cite{Kitazawa-Nomura-Okamoto}.  
We can eliminate the logarithmic corrections by taking the appropriate
average, for example, $(x^{p0} + 2 x^{p1}+x^{p3})$ as can be read from
Table 1.  We show the size dependence of 
$(x^{p0} + 2 x^{p1}+x^{p3})/4$ in Fig.4; the remaining size dependence
is mainly explained by the $x=4$ irrelevant field.  
From Table 1, in the neighborhood of the BKT transition, there appear
terms linear in the distance $t$ from the BKT line.  The ratios of 
$x^{p3}-x^{p1}$, $x^{m7}-x^{m3}$, $x^{m5}-x^{m2}$, $x^{m0}-x^{m3}$, 
are $-1/2:-1:1:-4/3$ close to the BKT line.  In Table 2, we show the
coefficients linear in $t$ of the eigenvalues corresponding with 
$x^{p3}-x^{p1}$, $x^{m7}-x^{m3}$, $x^{m5}-x^{m2}$, $x^{m0}-x^{m3}$.

\subsection{ S=2 XXZ chain}

Next we consider the S=2 XXZ spin chain,
described by the Hamiltonian:
\begin{equation}
	H= \sum (S^x_j S^x_{j+1}+ S^y_j S^y_{j+1} 
	+ \Delta S^z_j S^z_{j+1} + D ( S^z_j)^2).
	\label{eq:S=2-Hamiltonian}
\end{equation}
For the isotropic case ($\Delta=1$, $D=0$), 
several studies have been done in the relation to 
the Haldane's conjecture\cite{Haldane}
\cite{Hatano93,Deisz93,Sun95,Nishiyama95,Schollwock,Yamamoto95,Qin95,Qin97}. 
Although the estimated values of the Haldane gap are widely ranged, 
they are common on the existence and the smallness of it. 
For the smallness of the Haldane gap, we can expect that 
the transition points between the Haldane gap and the XY phases 
$1-\Delta_{c}$ ($D=0$) and $D_{c1}$ ($\Delta=1$) are very small, 
compared with the $S=1$ case\cite{Nightingale86,White92}. 
Here we only consider on the line $D=0$, and on the line $\Delta=0$ ($D>0$), 
to estimate the XY-Haldane and the XY-large $D$ transition points 
and to determine the universality class. 

First we consider on the $D=0$ line. 
Figure \ref{xcrss12} shows the excitation energies with $S^{z}_{T}=\pm 2$, 
$P=1$ under PBC, and with $S^{z}_{T}=0$ under TBC for $L=12$ systems. 
We can see a level crossing which corresponds to the transition point 
between the XY and the $S=2$ Haldane phases, as is expected from Table 1. 
The size dependence of this crossing point is shown in Fig. \ref{pnt}. 
The extrapolated value to the thermodynamic limit is $\Delta_{c}=0.966$.
The conformal anomaly number of this point is estimated as $c=1.16$. 
To check the universality class, we show the size dependence of 
the combination of the scaling dimension 
as $(x^{p0}+3x^{p1})/4$ in Fig. \ref{xdmnsn}. 
We extrapolate the $L\rightarrow \infty$ as 
$\Delta_{c}(L) = \Delta_{c}+a_{1}/L^{2}$ for $L=10,12$ and 
$\Delta_{c}(L) = \Delta_{c}+a_{1}/L^{2}+a_{2}/L^{4}$ for $L=8,10,12$, 
and the obtained values are $0.487$ and $0.491$ respectively. 

Next we show the result on the $\Delta=1$ ($D>0$) line. 
Figure \ref{crsslg} shows the excitation energies with $S^{z}_{T}=\pm 2$, 
$P=1$ under PBC, and with $S^{z}_{T}=0$ under 
TBC for $L=10$ systems. 
We can recognize the three regions as the $S=2$ Haldane phase $0<D<D_{c1}$, 
the massless XY phase $D_{c1}<D<D_{c2}$, and the large D phase $D_{c2}<D$. 
The size dependence of crossing points is shown 
in Fig. \ref{pntlb} and \ref{pntub}, and 
the estimated values are $D_{c1}=0.043$ and $D_{c2}=2.39$. 
The estimated conformal anomaly numbers are $c=1.16$ and $0.998$ for 
$D_{c1}$ and $D_{c2}$ respectively. 
The transition point between the $S=2$ Haldane gap and 
the $XY$ phases ($D_{c1}$) is consistent to the previously obtained values by 
Schollw\"ock and Jolic{\oe}ur \cite{Schollwock}($D_{c1}=0.004(2)$), 
but the transition point between the XY and the large-D phases ($D_{c2}$) 
deviates from their value ($D_{c2}\approx 3$).
Figure \ref{dldmnsn} and \ref{dudmnsn} shows 
the size dependence of $(x^{p0}+3x^{p1})/4$. 
The extrapolated values for $D_{c1}$ are $0.489$ for $L=10,12$ and 
$0.493$ for $L=8,10,12$. 
For $D_{c2}$ the values are $0.5005$ for $L=10,12$ and 
$0.5006$ for $L=8,10,12$. 

The conformal anomaly numbers and the scaling dimensions 
for the critical points 
$(\Delta,D)=(\Delta_{c},0)$ and $(1,D_{c1})$ 
somewhat deviate from the ideal value $c=1$ and $x=1/2$. 
We think that at these points the BKT transition points are 
very close to the 2D Ising transition point between the $S=2$ Haldane and 
the antiferromagnetic phases\cite{Schollwock}, 
so that the crossover effect is very large for these points. 
In fact, at the BKT transition point $(\Delta,D)=(1,D_{c2})$ 
where the 2D Ising transition point is far, 
these values are consistent to the ideal values. 

In the XY phase on the $\Delta=1$ line, we find that the energy levels 
with $S^{z}_{T}=0, U_{2\pi} P= U_{2\pi} T=1$ and with 
$S^{z}_{T}=0, U_{2\pi} P= U_{2\pi}T=-1$ cross 
two times (see Fig.\ref{fig:pntg}). These two points correspond 
to the Gaussian fixed points \cite{Kitazawa}, 
so that there are two Gaussian fixed lines in the whole phase diagram of 
(\ref{eq:S=2-Hamiltonian}). 
This may be the ``indirect'' evidence of the intermediate-$D$ 
phase, predicted by Oshikawa \cite{Oshikawa}. 
To see the intermediate-$D$ phase, Oshikawa, Yamanaka, 
and Miyashita\cite{Oshikawa-Yamanaka-Miyashita} 
studied the line $\Delta=1$ with the quantum Monte Carlo method. 
But we cannot find that phase, and 
we think that the intermediate-$D$ phase needs 
more large values of $\Delta$ ($>1$). 

Lastly we remark the following thing. 
From the large $S$ mapping onto the anisotropic nonlinear 
sigma model by Haldane\cite{Haldane}, $1-\Delta +D$ is proportional to the 
anisotropy of it. 
Numerical values $1-\Delta_{c}=0.034$ and $D_{c1}=0.043$ are comparable, 
thus it is consistent to Haldane's arguments.

\section{Conclusion}

Physical properties of the BKT transition, including 
the renormalization group properties and the $SU(2)$ 
or the $SU(2)/Z_2$ symmetry, 
have been well investigated in the field theory.  
However, the mapping from the various models to the field theoretical
models, such as the sine-Gordon model or the Wess-Zumino-Witten model, 
is not simple, not quantitatively correct.  
Fortunately, the symmetry structure and the sum rule at the BKT phase
transition point survive after the mapping.  
Therefore, using these properties, we can determine the BKT critical
point and the universality class.  

In numerical calculations, $SU(2)$ symmetry has been used to determine 
the BKT-type critical line in \cite{Nomura-Okamoto}, $SU(2)/Z_2$ symmetry has
been used in \cite{Nomura}, and the sum rule to eliminate 
logarithmic corrections in \cite{Ziman-Schulz}.
Recently, introducing the half magnetic charges by twisting boundary
condition, one of the authors has developed a method to determine the
Gaussian fixed line and its universality class for the non-integrable
models \cite{Kitazawa}.  
In this paper, using the twisted boundary condition, we explicitly 
relate the energy eigenvalue structure to the $SU(2)$ symmetry of the
BKT transition.  At the same time, we can improve the convergence of
the physical quantities to the thermodynamic limit, comparing with the
original level spectroscopy \cite{Nomura}.  

Our method is applicable not only to the quantum problems, but also 
to the classical models, treating the eigenvalue structure of 
the transfer matrix.

\section{Acknowledgement}

This work is partially supported by Grant-in-Aid for Scientific Research 
No. 09740308 from the Ministry of Education, Science and Culture, Japan.  
A. K. is supported by JSPS Research Fellowship for Young Scientists.  
The computation of this work has been done using the facilities of the
Supercomputer Center, Institute for Solid State Physics, 
University of Tokyo. 

\pagebreak

\appendix

\section{Free boson on the complex plane}

We shortly review the free boson theory on the complex plane \cite{Les-Houches}
and the relation between the half-odd magnetic monopole and twisted
boundary conditions \cite{Destri}.  
The equation of motion $\partial \bar{\partial} \phi=0$ for 
(\ref{eq:Gauss-Lagrangian}) allows the chiral decomposition 
\begin{equation}
	\phi (z,\bar{z}) = 
	\frac{\sqrt{K}}{2}(\varphi(z) + \bar{\varphi}(\bar{z})),
\end{equation}
where we introduce the two independent complex coordinates 
$z= x + iy, \bar{z}= x-iy$ and use the notation
$\partial = (\partial_x-i\partial_y)/2, 
\bar{\partial} = (\partial_x+ i\partial_y)/2$.  
Then the action can be written
\begin{equation}
	S = \int {\cal L} = \frac{1}{2\pi} \int \frac{ i dz \wedge d\bar{z}}{2} 
	\partial \varphi \bar{\partial} \bar{\varphi}.
	\label{eq:Gauss-Lagrangian2}
\end{equation}
The two point functions are
\begin{equation}
	\langle \varphi (z) \varphi(w) \rangle = - \log (z-w), 	\;
	\langle \bar{\varphi} (\bar{z}) \bar{\varphi}(\bar{w}) \rangle 
	= - \log (\bar{z}-\bar{w}).
	\label{eq:Green-function}
\end{equation}

{\em Chiral current: }
\newline
Here we introduce the U(1) chiral current as
\begin{equation}
	J(z) =i \partial \varphi (z).
\end{equation}
The chiral current has a leading short distance expansion
\begin{equation}
	J(z) J(w) = \frac{1}{(z-w)^2}+ \cdots,
	\label{eq:OPE-currents}
\end{equation}
inferred by taking two derivatives of (\ref{eq:Green-function}).  
We introduce the mode expansion of the current
\begin{equation}
	J(z) = \sum_{n} z^{-n-1} \alpha_n, 	\;
	\alpha_n = \oint \frac{d z}{2 \pi i} z^n J(z).
\end{equation}
Using the short distance expansion (\ref{eq:OPE-currents}) and 
the radial quantization, we obtain the commutation relation
\begin{equation}
	[\alpha_m, \alpha_n] = m \delta_{m+n,0},
\end{equation}
which means the U(1) current algebra.
Note that $\alpha_0$ is the conserved charge of the U(1) current.  

{\em Stress energy tensor: }
\newline
From the Noether theorem, the stress-energy tensor is written as 
\begin{equation}
	T(z) = \frac{1}{2} :J(z) J(z): .
\end{equation}
Using the short distance expansion (\ref{eq:OPE-currents}) and Wick's
theorem, we obtain OPE of the $T(z), T(w)$
\begin{equation}
	T(z)T(w) = \frac{1/2}{(z-w)^4} +\frac{2 T(w)}{(z-w)^2}
	+\frac{\partial T(w)}{z-w} + reg.,
\end{equation}
which means that this satisfies the $c=1$ Virasoro algebra. 
Similarly, OPE of the $T(z), J(w)$
\begin{equation}
	T(z)J(w) = \frac{J (w)}{(z-w)^2}
	+\frac{\partial J(w)}{z-w} + reg.
\end{equation}
means that $J(z)$ is the primary field with the conformal dimension (1,0).  
The mode expansion of $T(z)$ is
\begin{equation}
	T(z) = \sum_{n} z^{-n-2} L_n,
\end{equation}
where $L_n$ can be written
\begin{equation}
	L_n = \frac{1}{2} \sum_{m} :\alpha_{m+n} \alpha_{-m}:,
\end{equation}
especially
\begin{equation}
	L_0 = \frac{1}{2} \alpha_0^2 + 
	\sum_{n=1}^\infty :\alpha_{-n} \alpha_{n}:.
\end{equation}

{\em Vertex operator:}
\newline
Vertex operator is defined as
$
	:\exp(i \alpha \varphi(z)):
$.  
OPE of the vertex operator and the current is
\begin{equation}
	J(z):e^{i \alpha \varphi (w)} :
	= \frac{\alpha}{z-w}:e^{i \alpha \varphi (w)} : + reg..
	\label{eq:OPE-vertex-current}
\end{equation}
OPE of the vertex operator and the stress-energy tensor is
\begin{equation}
	T(z):e^{i \alpha \varphi (w)} :
	= \frac{\alpha^2/2}{(z-w)^2}:e^{i \alpha \varphi (w)} : 
	+ \frac{\partial_w:e^{i\alpha\varphi(w)}:}{z-w} + reg.,
\end{equation}
therefore the vertex operator is a primary field with 
conformal dimension $h=\alpha^2/2$.  

{\em Highest weight states: }
\newline
When we define the state
\begin{equation}
	| \alpha \rangle \equiv \lim_{w \rightarrow 0}
	:e^{i\alpha \varphi(w)}: | 0 \rangle,
\end{equation}
it is the highest weight state of the U(1) current algebra, 
\begin{equation}
	\alpha_0 | \alpha \rangle = \alpha |\alpha \rangle,	\;
	\alpha_n | \alpha \rangle = 0 \; (n>0),
\end{equation}
from eq. ({\ref{eq:OPE-vertex-current}).  
This is also the highest weight state of the Virasoro algebra
\begin{equation}
	L_0 | \alpha \rangle = \frac{\alpha^2}{2} |\alpha \rangle, 	\;
	L_n | \alpha \rangle = 0 \; (n>0).
	\label{eq:Highest-weight-state-Virasoro}
\end{equation}

{\em Compactification of the internal space:}
\newline
The compactification of the internal space $\phi \equiv \phi+2\pi/\sqrt{2}$ 
restricts the eigenvalues $\alpha,\bar{\alpha}$ of 
the $U(1)\times U(1)$ charges $\alpha_0,\bar{\alpha}_0$.  
First we note that 
\begin{eqnarray}
	\varphi(z) &=&q - i \alpha_0 \log z 
	+ i \sum_{n\neq 0} \frac{1}{n} z^{-n} \alpha_n, 	\nonumber	\\
	\bar{\varphi}(\bar{z}) &=& \bar{q} - i \bar{\alpha}_0 \log \bar{z} 
	+ i \sum_{n \neq 0} \frac{1}{n} \bar{z}^{-n} \bar{\alpha}_n, 
	\label{eq:Fubini-Veneziano}
\end{eqnarray}
where $q,\bar{q}$ are zero-modes and satisfy the commutation relation
\begin{equation}
	[q,\alpha_0]=[\bar{q},\bar{\alpha}_0]=i.
\end{equation}
Rotating around the origin $0$, there appears a phase factor
$\pm 2 \pi i$ in the function $\log (z)(\log (\bar{z}))$.  Thus, 
considering the periodic nature of $\phi$, we obtain
\begin{equation}
	\alpha-\bar{\alpha} = \sqrt{\frac{2}{K}} n, \; (n:\mbox{integer}),
	\label{eq:eigenvalue-alpha-}
\end{equation}
for the requirement of the uniqueness of the vertex operator under 
the change $\phi \equiv \phi+2\pi/\sqrt{2}$,
\begin{equation}
	\alpha+\bar{\alpha} = \sqrt{2K} m, \; (m:\mbox{integer}).
	\label{eq:eigenvalue-alpha+}
\end{equation}
In one word, eigenvalues for $\alpha_0,\bar{\alpha}_0$ are
\begin{equation}
	\alpha=\sqrt{\frac{K}{2}} m + \sqrt{\frac{1}{2K}} n,	\;
	\bar{\alpha}=\sqrt{\frac{K}{2}} m - \sqrt{\frac{1}{2K}} n.
\end{equation}
Since $L_0 \pm \bar{L}_0$ are the generators of dilatation and rotation, 
we identify $(\alpha^2 \pm \bar{\alpha}^2)/2$ as the scaling dimension 
and the conformal spin of the state $|\alpha,\bar{\alpha} \rangle$.  
Finally, by defining the dual field $\theta$ to $\phi$ as
\begin{equation}
	\theta = \frac{1}{2 \sqrt{K}} 
	(\varphi(z) - \bar{\varphi}(\bar{z})),
\end{equation}
$\theta$ has a periodicity $2\pi/\sqrt{2}$.  

{\em Twisted boundary conditions and Background charge: }
\newline
To any conformal operator $f(\alpha_0,\bar{\alpha}_0)$, we can
associate a twisted operator
\begin{eqnarray}
	f_{\Theta\bar{\Theta}} (\alpha_0,\bar{\alpha}_0) &=& 
	\exp (-i (\Theta q + \bar{\Theta} \bar{q})) f(\alpha_0,\bar{\alpha}_0)
	\exp (i (\Theta q + \bar{\Theta} \bar{q}))	\nonumber	\\
	&=& f(\alpha_0 + \Theta, \bar{\alpha}_0 + \bar{\Theta})
	\label{eq:Twisted-Operator}
\end{eqnarray}
Here we set 
\begin{equation}
	\Theta=\bar{\Theta} = \sqrt{\frac{K}{2}} \frac{\Phi}{2\pi}.
	\label{eq:Twisted-Angle}
\end{equation}
On the one hand, from eqs. (\ref{eq:Fubini-Veneziano}) this means 
\begin{equation}
	\phi \rightarrow \phi - i \frac{K}{\sqrt{2}} \frac{\Phi}{2\pi} \log |z|,	\;
	\theta \rightarrow \theta 
	- i \frac{1}{\sqrt{2}}\frac{\Phi}{2\pi} 
	(\frac{1}{2}\log \frac{z}{\bar{z}}),
\end{equation}
that is, there is a cut for $\theta$.  
Mapping from the plane to the cylinder by $w \equiv u +iv =(L/2\pi) \log z$, 
we obtain 
\begin{equation}
	\phi \rightarrow \phi - i \frac{K}{\sqrt{2}}\frac{\Phi}{L} u,	\;
	\theta \rightarrow \theta + \frac{1}{\sqrt{2}}\frac{\Phi}{L} v,
\end{equation}
that is, twisted boundary conditions for $\theta$ in the $v$
direction.  

On the other hand, the change (\ref{eq:Twisted-Operator}), 
(\ref{eq:Twisted-Angle}) with eqs. (\ref{eq:eigenvalue-alpha-}), 
(\ref{eq:eigenvalue-alpha+}) means the change of the magnetic charges
\begin{equation}
	m \rightarrow m + \frac{\Phi}{2\pi}, \; n \rightarrow n.
	\label{eq:Charges-change}
\end{equation}
This can be interpreted to set a magnetic monopole $\Phi/2\pi$ at the
origin $z=0$, and a magnetic monopole $-\Phi/2\pi$ at the infinity
$z=\infty$.  

{\em Discrete Symmetries}
\newline
For the Gaussian model (\ref{eq:Gauss-Lagrangian2}), 
besides continuous $U(1) \times U(1)$ symmetries, there are discrete symmetries
\begin{equation}
	T: (\varphi,\bar{\varphi}) \rightarrow (-\varphi,-\bar{\varphi}),\;
	C: (z,\varphi) \rightarrow (\bar{z},\bar{\varphi}).
\end{equation}

\section{Calculation of the renormalized scaling dimensions}

In this appendix, we calculate the correction 
of scaling dimensions in Table 1, 
up to the first order of $y_0,y_\phi$ in some degenerate cases. 
The derivation here is simpler than the original one 
\cite{Giamarchi-Schulz,Nomura}.  
Let us consider the following 1D quantum Hamiltonian 
\begin{equation}
  H = H_{0} + \frac{\lambda_{1}}{2\pi}\int_{0}^{L}dv {\cal{O}}_{1},
\end{equation}
where $H_{0}$ is a fixed point Hamiltonian, $L$ is the system size 
(in Table 1 $l$ is related with $L$ as $l= \log(L/\alpha)$), 
and ${\cal{O}}_{1}(={\cal{O}}_{1}^{\dagger})$ is a scaling operator whose 
scaling dimension is $x_{1}$. 
We set the short-range cutoff as $1$. 
According to Cardy \cite{Cardy86}, the size dependence of excitation energies 
up to the first order perturbation is given by 
\begin{equation}
  \Delta E_{n}(L) = \frac{2\pi}{L}\left( x_{n} +C_{n1n}\lambda_{1}(L)
  +\cdots \right) = \frac{2\pi}{L}x_{n}(L),
\label{finite}
\end{equation}
where $x_{n}$ is the scaling dimension of the operator ${\cal{O}}_{n}$, 
$C_{n1n}$ is the operator product expansion (OPE) coefficient of 
operators ${\cal{O}}_{n}$ and ${\cal{O}}_{1}$ as
\[
  {\cal{O}}_{1}(z,\bar{z}){\cal{O}}_{n}(0,0) 
  = C_{n1n}z^{-h_{1}}\bar{z}^{-h_{1}}{\cal{O}}_{n}(0,0) +\cdots,
\]
in which $h_{1}(=\bar{h}_{1})$ is the conformal weight of ${\cal{O}}_{1}$ 
($x_{1}=2h_{1}$). 
We used the notation 
\[
  \lambda_{1}(L) = \lambda_{1}\left(\frac{2\pi}{L}\right)^{x_{1}-2},
\]
which comes from the renormalization group equation. 

For the sine-Gordon model, we denote 
\[
  K = \frac{4}{m^{2}}\left( 1+\frac{1}{2}y_{0}\right),
\]
near $K=1$ ($m=2$) or $K=4$ ($m=1$).
We can rewrite the Lagrangian density as 
\begin{equation}
  {\cal{L}}(z,\bar{z}) =  {\cal{L}}_{0}(z,\bar{z})+{\cal{L}}_{I}(z,\bar{z}),
\end{equation}
where
\[
  {\cal{L}}_{0}(z,\bar{z}) = \frac{m^{2}}{8\pi}(\nabla\phi)^{2},
\]
and 
\begin{equation}
  {\cal{L}}_{I}(z,\bar{z}) = \frac{y_{0}}{4\pi}{\cal{M}}
    + \frac{y_{\phi}}{2\sqrt{2}\pi}\sqrt{2}\cos \sqrt{2} m\phi,
\end{equation}
and we set 
\begin{equation}
\lambda_{0} = \frac{y_{0}}{2},\hspace{5mm}
\lambda_{\phi}=\frac{y_{\phi}}{\sqrt{2}}.
\end{equation}

\subsection{$x^{p0}$ and $x^{p3}$}

We calculate $x^{p0}$ and $x^{p1}$ up to the first order of $y$'s. 
We denote operators 
$\sqrt{2}\cos m\phi/\sqrt{2}$ (whose scaling dimension is $x^{p0}$) and 
$\sqrt{2}\sin m\phi/\sqrt{2}$ (whose scaling dimension is $x^{p3}$) 
as ${\cal{O}}_{c}$ and ${\cal{O}}_{s}$ respectively. 
We have the following OPE's
\begin{equation}
{\cal{M}}(z,\bar{z}){\cal{O}}_{c,s}(0,0)
= \frac{1}{2}\frac{1}{|z|^{2}}{\cal{O}}_{c,s}
+\cdots,
\end{equation}
\begin{equation}
\sqrt{2}\cos\sqrt{2}m\phi(z,\bar{z}){\cal{O}}_{c,s}(0,0)
= \pm\frac{1}{\sqrt{2}}\frac{1}{|z|^{2}}{\cal{O}}_{c,s}(0,0)
+\cdots,
\end{equation}
where $+$ is for ${\cal{O}}_{c}$ and $-$ is for ${\cal{O}}_{s}$. 
From these OPE's and eq. (\ref{finite}), 
we can obtain the scaling dimensions,
\begin{equation}
  x^{p0}(L) = \frac{1}{2}+\frac{1}{2}\frac{y_{0}(L)}{2}
  +\frac{1}{\sqrt{2}}\frac{y_{\phi}(L)}{\sqrt{2}} 
\end{equation}
for $\sqrt{2}\cos m\phi/\sqrt{2}$, and 
\begin{equation}
  x^{p3}(L) = \frac{1}{2}+\frac{1}{2}\frac{y_{0}(L)}{2}
  -\frac{1}{\sqrt{2}}\frac{y_{\phi}(L)}{\sqrt{2}} 
\end{equation}
for $\sqrt{2}\sin m\phi/\sqrt{2}$. 
Setting $y_{\phi}=y_{0}(1+t)$, we have the scaling dimension described in 
Table 1. These are consistent with the results obtained by Giamarchi and 
Schulz \cite{Giamarchi-Schulz}.  

Similar calculation can be applied to $x^{m5,m6,m7,m8}$.

\subsection{Marginal operators}

For the BKT transition, 
there exists a hybridization between the marginal field ${\cal{M}}$ 
and the operator $\sqrt{2}\cos\sqrt{2} m\phi$ ($m=2$ for $K=1$, and 
$m=1$ for $K=4$) \cite{Nomura}.  
These two operators have the same scaling dimension and symmetries 
at $K=1$ or $K=4$. 
Treating ${\cal{L}}_{I}$ as the perturbation term, we have the following OPE's
\begin{eqnarray}
  {\cal{L}}_{I}(z,\bar{z}){\cal{M}}(0,0) &=& 
  \frac{y_{\phi}\sqrt{2}}{2\pi |z|^{2}}\sqrt{2}\cos\sqrt{2}m\phi(0,0) 
  +\cdots, \\
  {\cal{L}}_{I}(z,\bar{z})\sqrt{2}\cos\sqrt{2}m\phi(0,0) &=& 
  \frac{y_{0}}{2\pi |z|^{2}}\sqrt{2}\cos\sqrt{2} m\phi(0,0) 
  + \frac{y_{\phi}\sqrt{2}}{2\pi |z|^{2}} {\cal{M}}(0,0) +  \cdots.
  \nonumber
\end{eqnarray}
Setting $y_{\phi}=y_{0}(1+t)$, and diagonalizing these equations, 
we have the orthogonal operators,
\begin{equation}
  \sqrt{\frac{2}{3}}\left(1-\frac{1}{9}t\right){\cal{M}}
  + \sqrt{\frac{1}{3}}\left(1+\frac{2}{9}t\right)\sqrt{2}\cos\sqrt{2}m\phi
\end{equation}
with the scaling dimension (up to the first order of $y_{0}$ and $t$)
\[
  x^{m0}(L) = 2-y_{0}(L)\left(1+\frac{4}{3}t\right), 
\]
and 
\begin{equation}
  \sqrt{\frac{2}{3}}\left( 1 - \frac{1}{9}t \right)\sqrt{2}\cos\sqrt{2}m\phi
  -\sqrt{\frac{1}{3}}\left( 1+\frac{2}{9}t\right){\cal{M}}
\end{equation}
with the scaling dimension
\[
  x^{m1} = 2 + 2y_{0}(L)\left(1+\frac{2}{3}t\right).
\]
These scaling dimensions are consistent with the results obtained by 
one of the authors \cite{Nomura}. 

Scaling dimensions of remaining operators in Table 1 does not split.

\section{Symmetry in twisted boundary conditions}

In eq. ({\ref{eq:TBC2}), we showed the unitary transformation of TBC 
for spin operators.  In this appendix we discuss the unitary
transformation of other operators $T_R, P,T$.  They are not well defined
except for the special $\Phi$, in contrast to the spin operators.  

\subsection{Translation operator}

The unitary operator $U_\Phi$ of the twisted boundary condition 
(\ref{eq:TBC2}) is transformed with the translation as
\begin{eqnarray}
	T_R U_\Phi [T_R]^{-1} &=& \exp \left( \frac{i \Phi}{L}
	\sum_{j=1}^L (j-1/2) S^z_{j+1}\right)	\nonumber	\\
	&=& U_{\Phi} \exp \left( -\frac{i\Phi}{L}S^z_T \right)
	\exp(i\Phi S^z_1),
	\label{eq:Translation-Unitary}
\end{eqnarray}
therefore, we obtain
\begin{equation}
	U_\Phi T_R [U_\Phi]^{-1} = \exp(i\Phi S^z_1) T^t_R,
\end{equation}
where we introduce the operator $T_R^t \equiv \exp (i \Phi S^z_T/L) T_R$.  
The operator $\exp(i\Phi S^z_1)$ is not well defined except for
$\Phi=2\pi l$($l$:integer)
\begin{equation}
	U_{2\pi l} T_R [U_{2\pi l}]^{-1} = (-1)^{2Sl} T^t_R.
\end{equation}
In the notation $T_R^t$, the periodicity of the energy 
and the momentum eigenvalue under 
$\Phi \rightarrow \Phi +2\pi$($S$: integer) or 
$\Phi \rightarrow \Phi +4\pi$($S$: half-odd-integer) 
\cite{Sutherland-Shastry,Fukui-Kawakami} becomes apparent,
in contrast to \cite{Kolb,Fath-Solyom}.  
Note that when we define the wavenumber $q' \equiv q+\Phi S^z_T/L$ as 
\begin{equation}
	T_R^t|\psi(q,S^z_T,\Phi) \rangle
	= \exp (i (q +\Phi S^z_T/L)) |\psi(q,S^z_T,\Phi) \rangle,
\end{equation}
$Lq'/2\pi$ is not integer in general, though it corresponds to 
the eigenvalue $mn + n\Phi/2\pi$ of the conformal spin operator 
$L_0-\bar{L}_0$ under TBC (see eqs. 
(\ref{eq:Highest-weight-state-Virasoro}), (\ref{eq:eigenvalue-alpha-}), 
(\ref{eq:eigenvalue-alpha+}), (\ref{eq:Charges-change}) ), 
and $T_R^t$ has 
a more natural symmetry structure than $T_R$, as will be seen later.  

\subsection{Discrete symmetries $P,T$}

The unitary operator $U_\Phi$ is transformed with the parity as
\begin{eqnarray}
	P U_\Phi P &=& \exp 
	\left(\frac{i\Phi}{L} \sum_{j=1}^L (j-1/2) S^z_{L+1-j} \right)
	\nonumber	\\
	&=& U_{-\Phi} \exp (i\Phi S^z_T),
\end{eqnarray}
therefore, we obtain
\begin{equation}
	U_\Phi P [U_\Phi]^{-1} = U_{2\Phi}P \exp(-i\Phi S^z_T),
\end{equation}
The operator $\exp(i\Phi S^z_T)$ is not well defined except for
$\Phi=\pi l$($l$:integer)
\begin{equation}
	U_{\pi l} P [U_{\pi l}]^{-1} = (-1)^{S^z_T l} U_{2\pi l} P,
\end{equation}
where we use that $S^z_T$ is integer.  In this case, we obtain
\begin{equation}
	(U_{2\pi l} P)^2 =1.
\end{equation}
Similarly, we obtain
\begin{equation}
	U_\Phi T [U_\Phi]^{-1} = U_{2\Phi} T,
\end{equation}
and 
\begin{equation}
	(U_{2\Phi} T)^2 = 1.
\end{equation}

For the twist angle $\Phi=\pi l$, Hamiltonian 
(\ref{eq:S=1-Twisted-Hamiltonian}) commutes with the
operators $U_{2\pi l} P, U_{2\pi l} T$.  Moreover, these operators forms
$Z_2$ group, that is,
the eigenvalue of $U_{2\pi}P,U_{2\pi}T$ is $\pm 1$.  

Next for the twist angle $\Phi =\pi l$, 
we discuss the relation between the operator $U_{2\pi l} P$ 
and the translation operator $T_R^t$. 
Using (\ref{eq:Translation-Unitary}), we obtain
\begin{eqnarray}
	T_R^t U_{2\pi l} P 	T_R^t
	&=& \exp ( \frac{2\pi i l}{L} S^z_T) T_R U_{2\pi l} [T_R]^{-1} 
	T_R P T_R
	\nonumber	\\
	&=& \exp (2\pi i S^z_1 l) U_{2\pi l} P.
\end{eqnarray}
Since the operator $\exp (2\pi i S^z_1)$ is $\pm I$ for $S$ integer or
half-integer, this means
\begin{equation}
	T_R^t  (U_{2\pi l } P) = (-1)^{2Sl}  (U_{2\pi} P) (T_R^t )^{-1},
\end{equation}
that is, in the momentum space for the even-integer $2Sl$, the spectrum is
symmetric with respect to $q=0,\pi$, whereas for the odd-integer $2Sl$, the
spectrum is symmetric with respect to $q= \pm \pi/2$.  
For the half-odd integer $S$ case, under TBC ($\Phi=\pi$) 
the ground state ($q=0$) 
is exactly degenerate with a state $q=\pi$ \cite{Kolb}, which is
related to the Gaussian transition \cite{Kitazawa-Nomura}.

\subsection{VBS states}

As an application of the previous subsections, 
we discuss the generalized $Z_2 \times Z_2$ symmetries or valence bond
solid (VBS) \cite{Oshikawa} under the twisted boundary condition.  
It was shown that under the twisted boundary conditions the quantum
numbers $P,T$ are the good quantum numbers characterizing the generalized
$Z_2 \times Z_2$ symmetries \cite{Kitazawa-Nomura}.  Here we
generalize these results for the twisted boundary conditions with
the translational invariant case.  

The spin variable can be represented by the Schwinger bosons as follows 
\begin{equation}
	S^z_j = \frac{1}{2}(a^+_j a_j - b^+_j b_j), \;
	S^+_j = a^+_j b_j,	\;
	S^-_j = a_j b^+_j,	\;
\end{equation}
with the constraint that the boson occupation number at each site 
$a^+_j a_j + b^+_j b_j$ is $2S$.  

The VBS states with the TBC can be written as
\begin{eqnarray}
	|S,M,TBC(\Phi=\pi) \rangle &=&
	(a^+_L b^+_1 e^{-i\pi/2L} - b^+_L a^+_1 e^{i\pi/2L})^{S-M}
	\nonumber	\\
	&\times&	\prod_{j=1}^{L/2-1}
	(a^+_{2j-1} b^+_{2j} e^{-i\pi/2L} - b^+_{2j-1} a^+_{2j} e^{i\pi/2L})^{S+M}
	\nonumber	\\	
	&\times& (a^+_{2j} b^+_{2j+1} e^{-i\pi/2L} - b^+_{2j} a^+_{2j+1} e^{i\pi/2L})^{S-M}
	\nonumber	\\
	&\times& (a^+_{L-1} b^+_L e^{-i\pi/2L} - b^+_{L-1} a^+_L e^{i\pi/2L})^{S+M}
	| 0 \rangle
\end{eqnarray}
where $M$ is an integer for the integer $S$, or a half-odd integer fot
the half-odd integer $S$ 
(here we include bond-alternating cases).  
First we make a parity transformation for the VBS state
\begin{equation}
	P| S, M, TBC(\Phi=\pi) \rangle
	= (-1)^{SL} | S,M,TBC(\Phi=-\pi) \rangle
\end{equation}
where we use $P|0 \rangle = | 0 \rangle$.  Then twisting with
$U_{2\pi}$, we obtain 
\begin{equation}
	U_{2\pi} P| S, M, TBC(\Phi=\pi) \rangle
	= (-1)^{SL-S+M} | S,M,TBC(\Phi=\pi) \rangle
\end{equation}
where we use $U_{2\pi}|0 \rangle = | 0 \rangle$,  
and 
$U_{2\pi} a_L^+ b_1^+ U_{2\pi}^{-1} = a_L^+ b_1^+ \exp(2\pi
i(L-1)/2L)$.  
The same discussion applies for $U_{2\pi}T$.  Therefore, 
each M-VBS states is characterized by the discrete quantum numbers 
$U_{2\pi} P =U_{2\pi} T = (-1)^{SL-S+M}$.  

Similarly we can classify the intermediate large D phase 
with the discrete quantum number $U_{2\pi} P,U_{2\pi} T$ 
under TBC.  

\pagebreak

\pagebreak

\input{table1.tex}

\input{table2.tex}

\input{fig.tex}

\end{document}

%% file: table1.tex
\begin{table}

\small

\caption{
Operator content of the sine-Gordon model at $K=1$ and $K=4$.
Here we consider the BKT transition of the $y_\phi(l) =y_0(l)$ branch, 
and we denote the deviation $t$ from the BKT critical line as 
$y_\phi (l) =y_0(l)(1+t)$
(for the $y_\phi (l) = - y_0(l)(1+t)$ branch, 
the role of operators $x^{p0} \leftrightarrow x^{p3}, \;
x^{m5,m6} \leftrightarrow x^{m7,m8}$ interchanges).  
*:Strictly speaking, these operators ($x^{m0},x^{m1}$) 
are hybridized under renormalization. 
$^{\dagger}$ : P,T should be interpreted as $U_{2\pi}T, U_{2\pi}P$ 
under TBC.  
}

\bigskip

\begin{tabular}{ccccc}
\hline
\multicolumn{1}{c}{}	&
\multicolumn{1}{c}{Operator in}	&
\multicolumn{1}{c}{Operator in}	&
\multicolumn{1}{c}{Renormalized}&
\multicolumn{1}{c}{Dicrete}	\\
\multicolumn{1}{c}{}	&
\multicolumn{1}{c}{s-G model (K=1)}	&
\multicolumn{1}{c}{s-G model (K=4)}	&
\multicolumn{1}{c}{scaling dimension}		&
\multicolumn{1}{c}{symmetries$^{\dagger}$}		\\

\hline

$x^{p0}$	& $O_{1,0}+O_{-1,0}$	& $O_{1/2,0}+O_{-1/2,0}$(TBC)	
& $\frac{1}{2}+\frac{3}{4}y_0(l)(1+\frac{2}{3}t)$	&	$P=T=1$ 	\\

$x^{p1,p2}$	& $O_{0, \pm1}$	& $O_{0,\pm2}$
& $\frac{1}{2}-\frac{1}{4}y_0(l)$ 	&	$P=1$	\\

\bigskip

$x^{p3}$	& $O_{1,0}-O_{-1,0}$	& $O_{1/2,0}-O_{-1/2,0}$(TBC)	
& $\frac{1}{2}-\frac{1}{4}y_0(l)(1+2 t)$ 	&	$P=T=-1$	\\

$x^{c0}$	& $\frac{2i}{\sqrt{K}} \partial \phi$	
& $\frac{2i}{\sqrt{K}}\partial \phi$
& $1$ 	&	\\

\bigskip

$x^{c1,c2}$	& $O_{\pm 1,\pm 1}$	& $O_{\pm 1/2, \pm 2}$(TBC)
& $1$ 	&	\\

$x^{m0}$*	& $-\frac{4}{K}\partial \phi \bar{\partial} \phi$	
& $-\frac{4}{K}\partial \phi \bar{\partial} \phi$
& $2 - y_0(l)(1+\frac{4}{3}t) $ 	&	$P=T=1$	\\

$x^{m1}$*	& $O_{2,0}+O_{-2,0}$	& $O_{1,0}+O_{-1,0}$ 
& $2 + 2y_0(l)(1+\frac{2}{3}t) $ 	&	$P=T=1$	\\

$x^{m2}$	& $O_{2,0}-O_{-2,0}$	& $O_{1,0}-O_{-1,0}$ 
& $2 + y_0(l) $ 	&	$P=T=-1$	\\

$x^{m3,m4}$	& $O_{0,\pm 2}$	& $O_{0,\pm 4} $ 
& $2 - y_0(l) $ 	&	$P=1$	\\

$x^{m5,m6}$	& $\bar{\partial} O_{\pm 1,\pm 1}+\partial O_{\mp 1,\pm 1}$	
& $\bar{\partial} O_{\pm 1/2, \pm 2}+\partial O_{\mp 1/2, \pm 2}$(TBC)
& $2 +y_0(l)(1+t)$ 	&	$P=1$	\\

$x^{m7,m8}$	& $\bar{\partial} O_{\pm 1,\pm 1}-\partial O_{\mp 1,\pm 1}$	
& $\bar{\partial} O_{\pm 1/2, \pm 2}-\partial O_{\mp 1/2, \pm 2}$(TBC)
& $2 -y_0(l)(1+t)$ 	&	$P=-1$	\\

\hline

\end{tabular}

\end{table}

%% file: table2.tex
\begin{table}[ht]
\begin{center}
\begin{tabular}{ccc}\hline
 excitations &$\mbox{d}\Delta E/\mbox{d}\Delta|_{\Delta=0}$&expected ratio \\
\hline
  p1,p3  &  0.50426315 & 1/2  \\
  m3,m0  &  1.23831439 & 4/3 \\
  m7,m0  &  0.30949032 & 1/3 \\
  m3,m7  &  0.92882406 & 1\\
  m5,m2  &  0.93899885 & 1\\
\hline
\end{tabular}
\end{center}
\caption{$ \mbox{d} \Delta E/\mbox{d} \Delta|_{\Delta=0}$ for $L=16$ system.}
\end{table}

%% file: fig.tex
\begin{figure}[ht]
\begin{center}
\leavevmode
\epsfxsize=4in \epsfbox{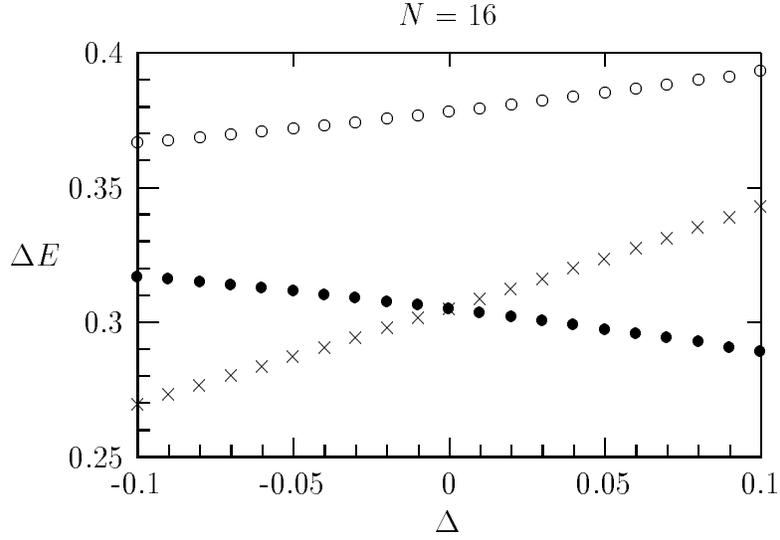}
\end{center}
\caption{Excitation energies of $S^{z}_T =\pm 2, q=0, P=1$ state 
with PBC ($\times$), 
and of $S^{z}_{tot}=0, q=0$ states with TBC 
($\bullet: U_{2\pi}P = U_{2\pi}T=-1$ and $\circ : U_{2\pi}P=U_{2\pi}T=1$). 
}
\label{crss16}
\end{figure}

\begin{figure}
\begin{center}
\leavevmode
\epsfxsize=4in \epsfbox{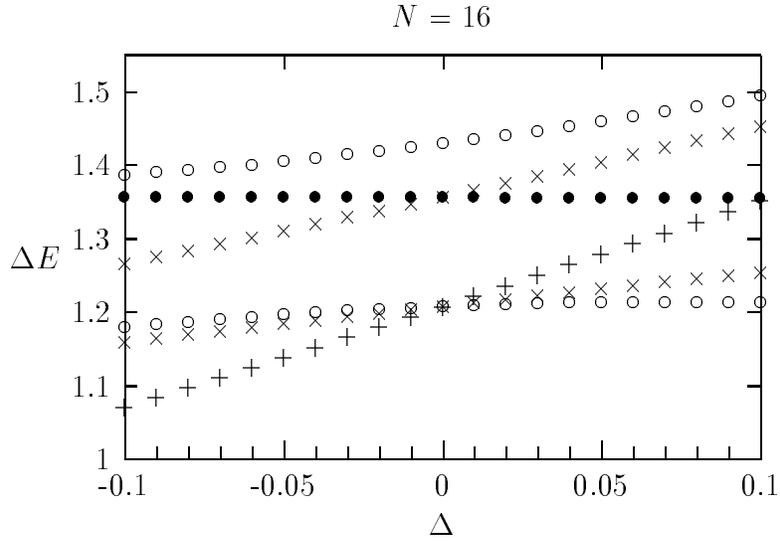}
\end{center}
\caption{Excitation energies of $S^{z}_T=0, q=0, P=T=1$ states 
with PBC ($\circ$), 
of $S^{z}_T=0, q=0, P=T=-1$ state with PBC ($\bullet$),  
of $S^{z}_T = \pm 2, q=0$ states with TBC ($\times$), 
and of $S^{z}_T=\pm 4, q=0, P=1$ state with PBC ($+$), 
}
\label{mrgnl16}
\end{figure}

\begin{figure}
\begin{center}
\leavevmode
\epsfxsize=4in \epsfbox{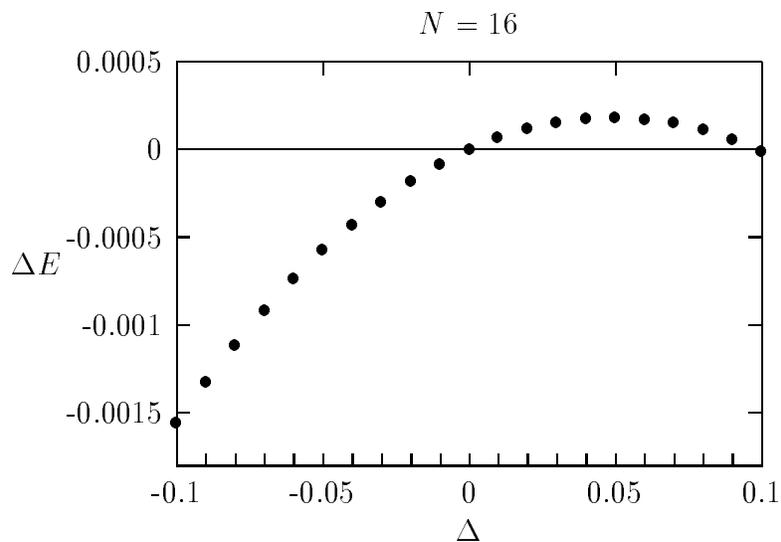}
\end{center}
\caption{Energy difference $E^{c0}-E^{c1}$ for $L=16$ systems.}
\label{current16}
\end{figure}
\vspace{2cm}

\begin{figure}
\begin{center}
\leavevmode
\epsfxsize=4in \epsfbox{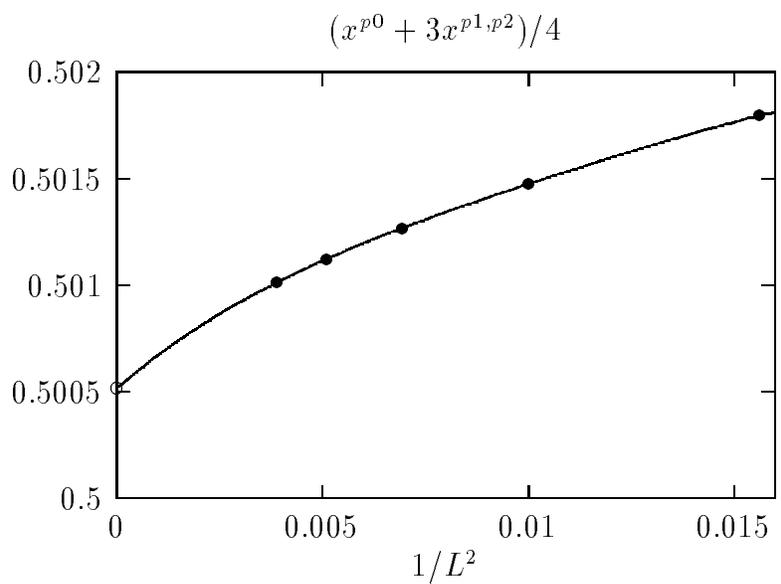}
\end{center}
\caption{Size dependence of the averaged scaling dimension 
$(x^{p0}+3x^{p1,p2})/3$.}
\label{prmry}
\end{figure}

\begin{figure}[ht]
\begin{center}
\epsfxsize=4in 
\leavevmode\epsfbox{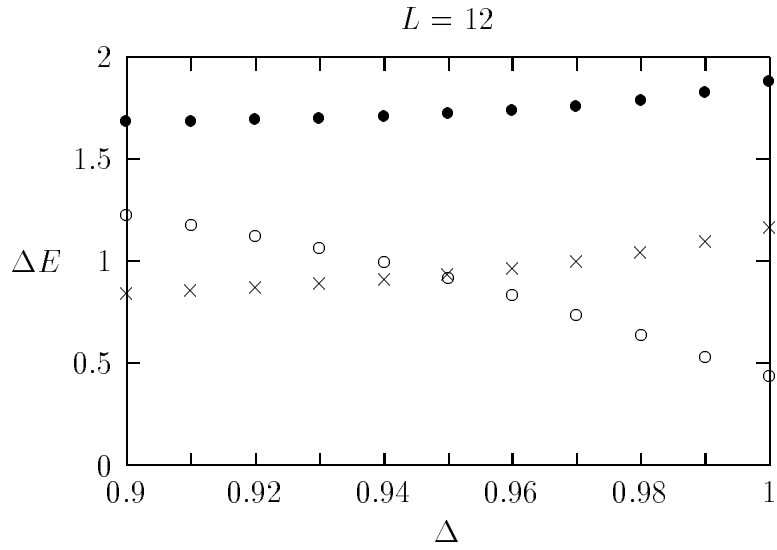}
\end{center}
\caption{Excitation energies of $L=12$, $D=0$ near 
the XY-Haldane transition point. 
$\times$'s are $S^{z}_{T}=\pm 2$, $q=0$, $P=1$ excitations under PBC, 
$\circ$'s are $S^{z}_{T}=0, U_{2\pi}P=U_{2\pi}T=1$ 
excitations under TBC, and 
$\bullet$'s are $S^{z}_{T}=0, U_{2\pi}P=U_{2\pi}T=-1$ under TBC.}
\label{xcrss12}
\end{figure}

\begin{figure}[ht]
\begin{center}
\epsfxsize=4in 
\leavevmode\epsfbox{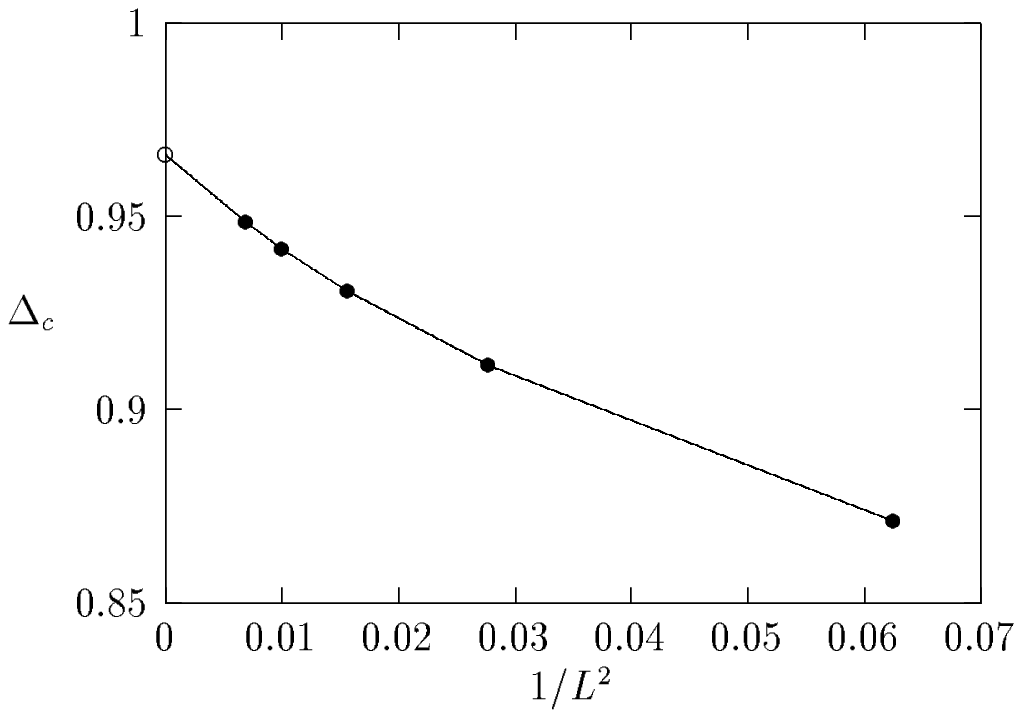}
\end{center}
\caption{
Size dependence of the crossing point. The extrapolated value 
is $\Delta_{c}=0.966$, 
which is the transition point between the $S=2$ Haldane gap and the 
XY phases.
}
\label{pnt}
\end{figure}

\begin{figure}[ht]
\begin{center}
\epsfxsize=4in 
\leavevmode\epsfbox{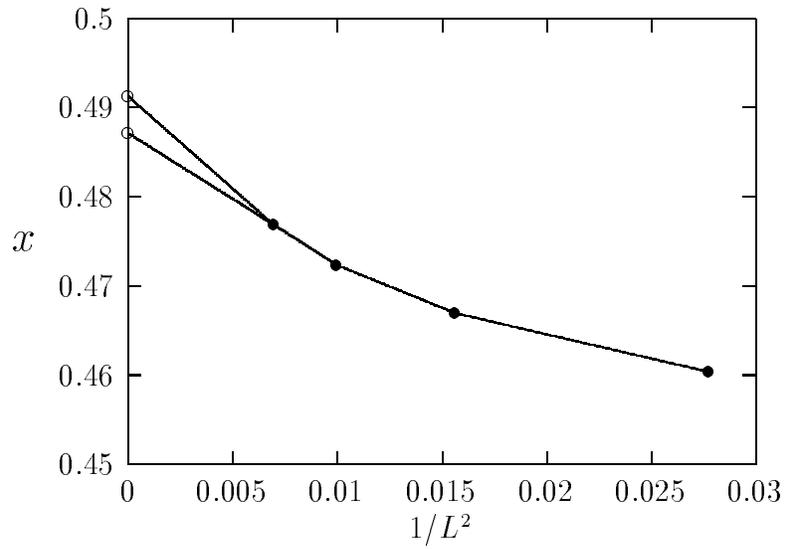}
\end{center}
\caption{Size dependence of the averaged scaling dimension 
$(x^{p0}+3x^{p1})/4$ for $\Delta_{c}$}
\label{xdmnsn}
\end{figure}

\begin{figure}[ht]
\begin{center}
\leavevmode
\epsfxsize=4in \epsfbox{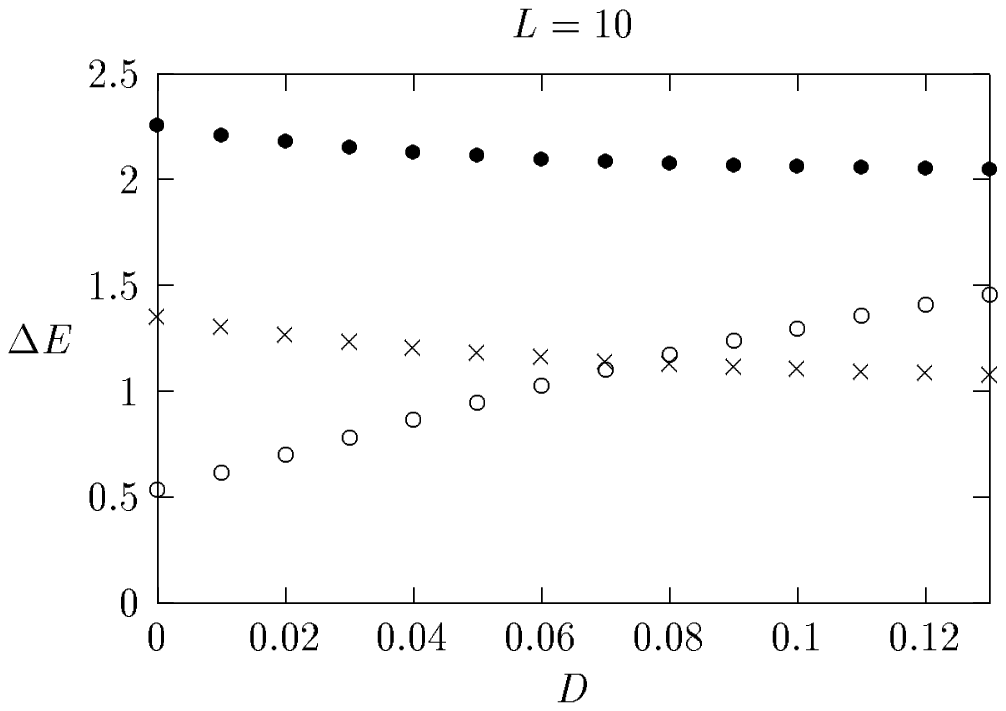}
\vspace{1cm}

\leavevmode
\epsfxsize=4in \epsfbox{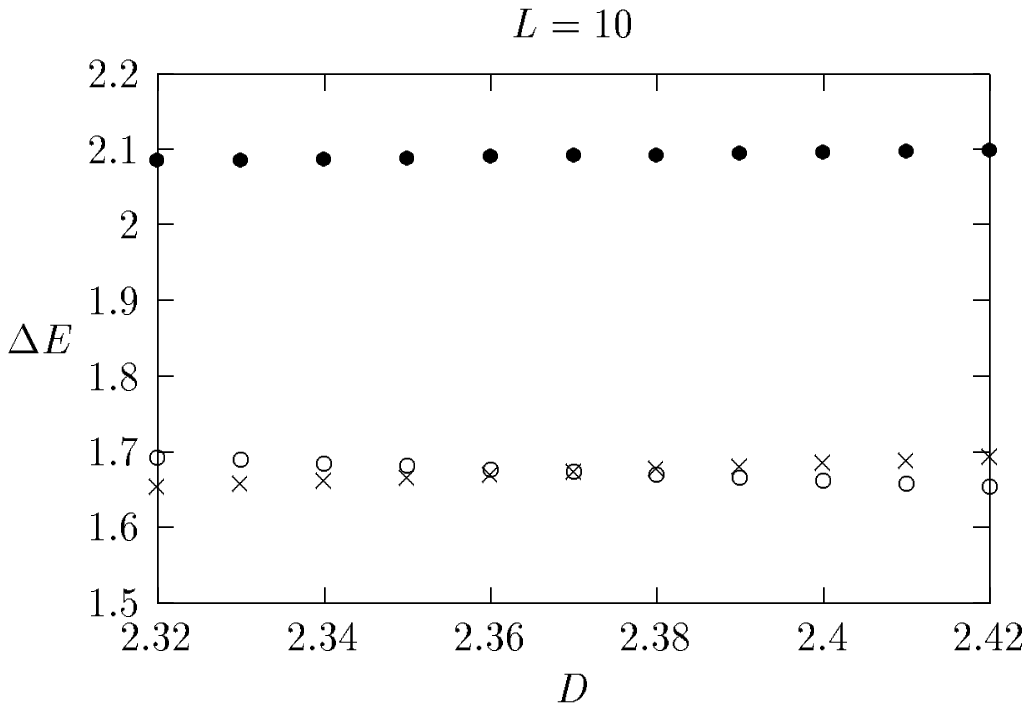}
\end{center}
\caption{Excitation energies of $L=10$, $\Delta=1$ near 
the Haldane-XY and XY-large $D$ transition points. 
$\times$'s are $S^{z}_{T}=\pm 2$, $q=0$, $P=1$ excitations under PBC, 
$\circ$'s are $S^{z}_{T}=0, U_{2\pi}P=U_{2\pi}T=1$ 
excitations under TBC, and 
$\bullet$'s are $S^{z}_{T}=0, U_{2\pi}P=U_{2\pi}T=-1$ under TBC.}
\label{crsslg}
\end{figure}

\begin{figure}[ht]
\begin{center}
\leavevmode
\epsfxsize=4in \epsfbox{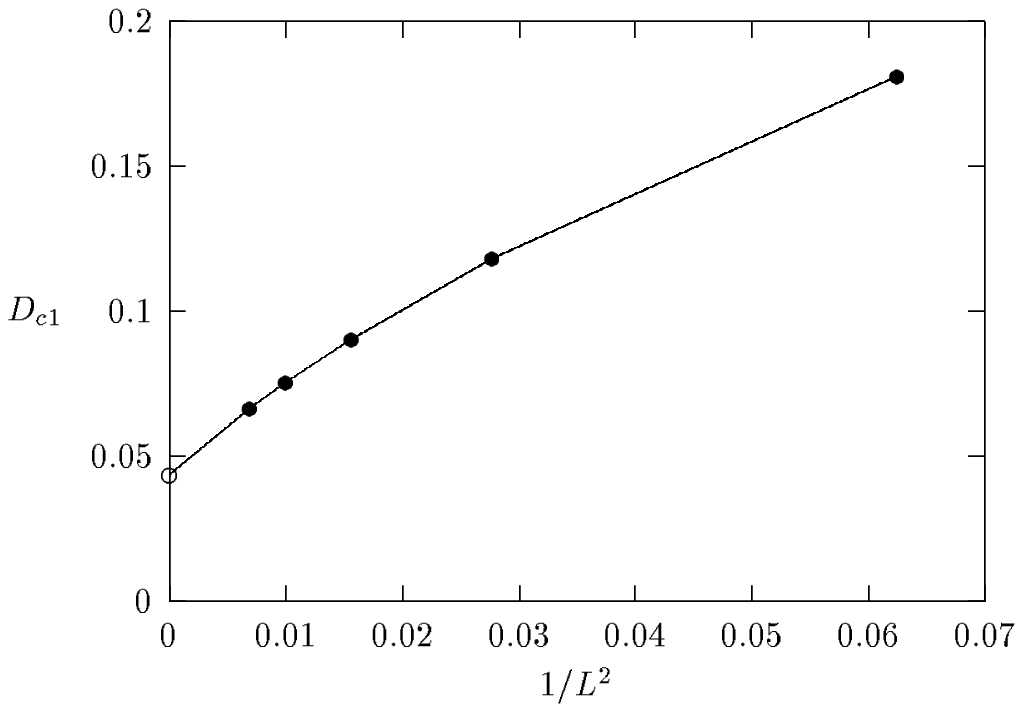}
\end{center}
\caption{Size dependence of the crossing point. The extrapolated value 
is $D_{c1}=0.043$, 
which is the transition point between the $S=2$ Haldane gap and the 
XY phases.}
\label{pntlb}
\vspace{1cm}

\begin{center}
\leavevmode
\epsfxsize=4in \epsfbox{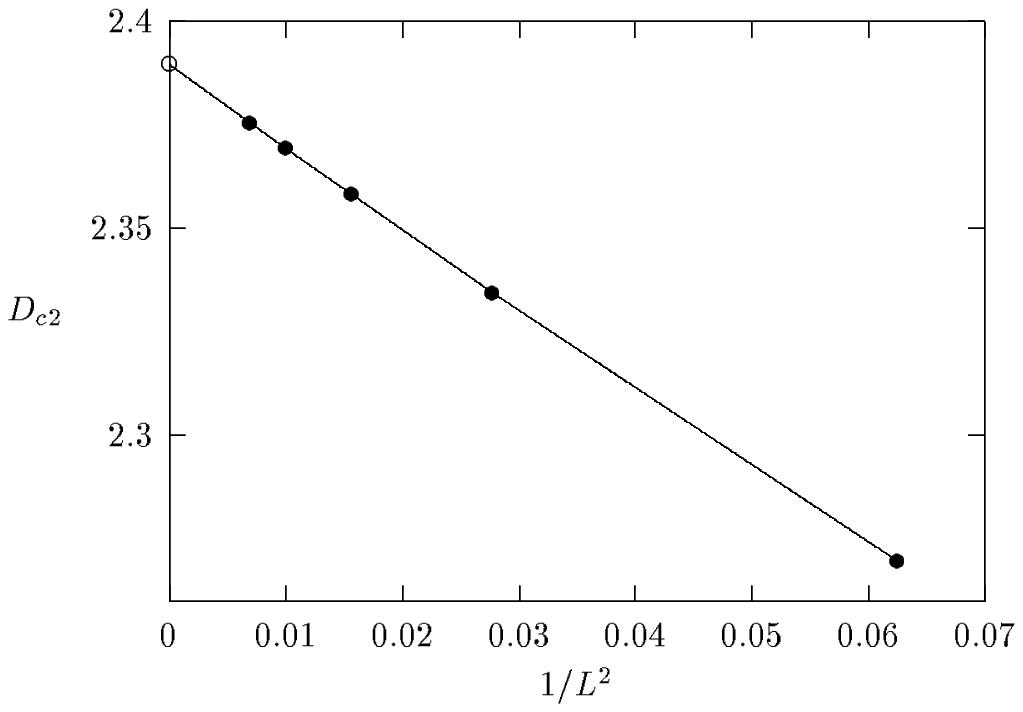}
\end{center}
\caption{Size dependence of the crossing point. The extrapolated value 
is $D_{c2}=2.39$, which is the transition point between the XY and the 
large $D$ phases.}
\label{pntub}
\end{figure}

\begin{figure}
\begin{center}
\leavevmode
\epsfxsize=4in \epsfbox{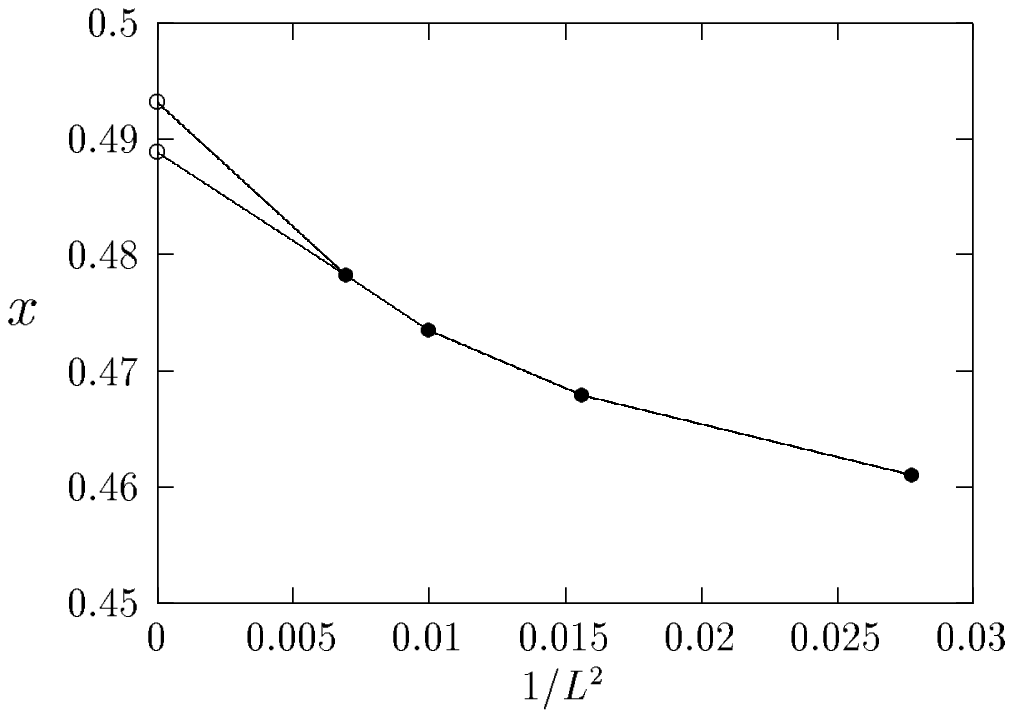}
\end{center}
\caption{Size dependence of the averaged scaling dimension for $D_{c1}$.}
\label{dldmnsn}
\end{figure}

\begin{figure}
\begin{center}
\leavevmode
\epsfxsize=4in \epsfbox{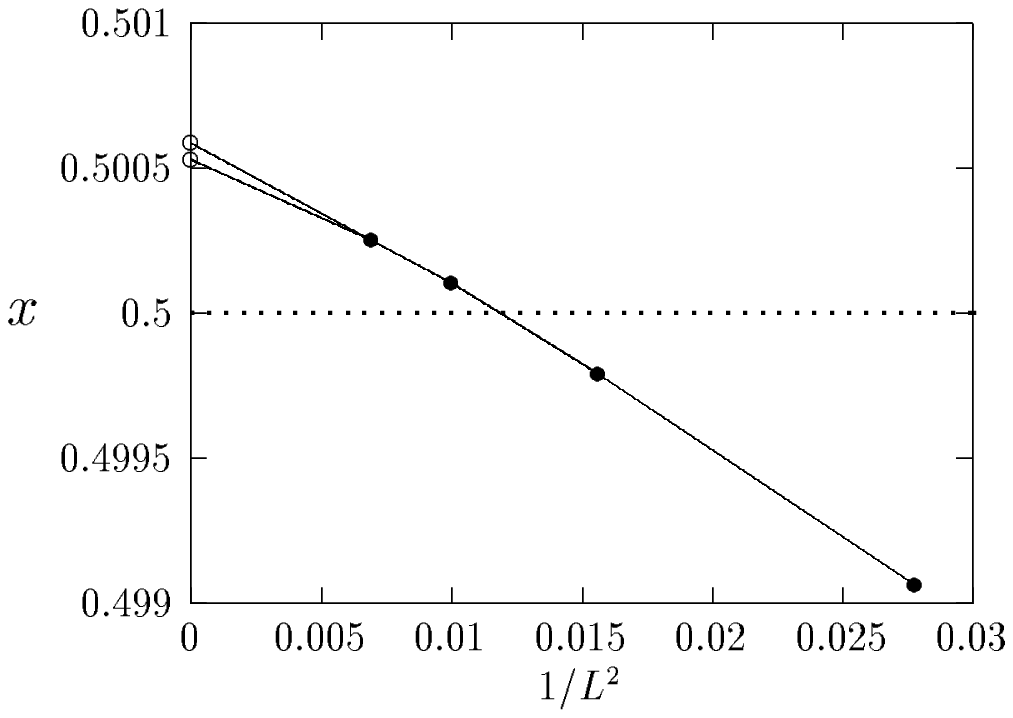}
\end{center}
\caption{Size dependence of the averaged scaling dimension for $D_{c2}$}
\label{dudmnsn}
\end{figure}

\begin{figure}
\begin{center}
\leavevmode
\epsfxsize=4in \epsfbox{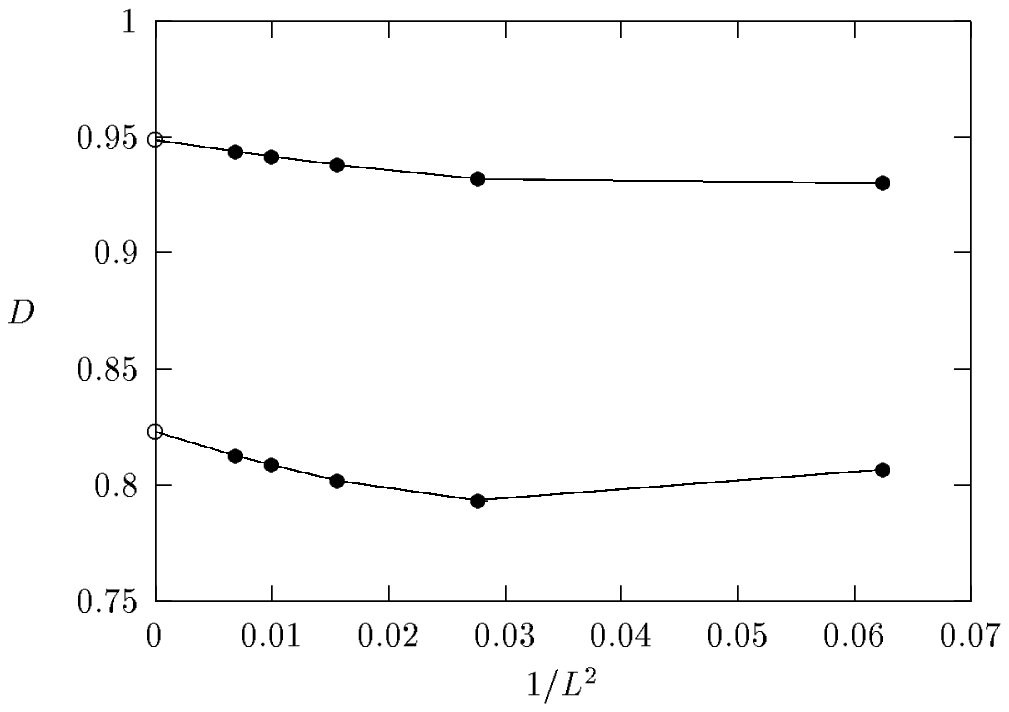}
\end{center}
\caption{Size dependence of the Gaussian fixed points in XY phase 
on the $\Delta=1$ line}
\label{fig:pntg}
\end{figure}

%% file: paper.bbl
\begin{thebibliography}{99}
\bibitem{Berezinskii} V. L. Berezinskii: Zh. Eksp. Teor. Fiz. {\bf 61} 
	(1971) 1144 (Sov. Phys.-JETP {\bf 34} (1972) 610).
\bibitem{Kosterlitz-T} J. M. Kosterlitz and D. J. Thouless: 
	J. Phys. C {\bf 6} (1973) 1181.
\bibitem{Kosterlitz} J. M. Kosterlitz: J. Phys. C {\bf 7} (1974) 1046.
\bibitem{Giamarchi} T. Giamarchi: Phys. Rev. B 44, (1991) 2905; 
	Phys. Rev. B 46 (1992) 342.
\bibitem{Gupta} For example, R. Gupta, J. DeLapp, G. G. Batrouni, 
	G. C. Fox, C. F. Baillie, and J. Apostolakis: 
	Phys. Rev. Lett. {\bf 61} (1988) 1996.
\bibitem{Seiler} E. Seiler, I. O. Stamatescu, A. Patrascioiu, 
	and V. Linke: Nucl. Phys. B {\bf 305} [FS23] (1988) 623.
\bibitem{Edwards} R. G. Edwards, J. Goodman, and A. D. Sokal: 
	Nucl. Phys. B {\bf 354} (1991) 289.
\bibitem{Solyom} J. S\'olyom and T. A. L. Ziman: Phys. Rev. B 
	{\bf 30} (1984) 3980. 
\bibitem{Nomura} K. Nomura: J. Phys. A {\bf 28} (1995) 5451.
\bibitem{Halpern} M. B. Halpern: Phys Rev. D {\bf 12} (1975) 1684;
	Phys. Rev. D {\bf 13} (1976) 337.
\bibitem{Banks} T. Banks, D. Horn, and H. Neuberger: Nucl. Phys. B 
	 {\bf 108} (1976) 119.
\bibitem{Ginsparg} P. Ginsparg: Nucl. Phys. B {\bf 295}[FS21] (1988) 153.
\bibitem{Blote-Cardy-Nightingale} H. W. J. Bl\"ote, J. L. Cardy, 
	and M. P. Nightingale: 
	Phys. Rev. Lett. {\bf 56} (1986) 742.
\bibitem{Destri} C. Destri and H. J. de Vega: Phys. Lett. B, {\bf 223} 
	(1989) 365.
\bibitem{Kitazawa} A. Kitazawa: J. Phys. A {\bf 30} (1997) L285.
\bibitem{Alcaraz-Barber-Batchelor} F. C. Alcaraz, M. N. Barber, 
	and M. Batchelor: Phys. Rev. Lett. {\bf 58} (1987) 771; 
	Ann, Phys. {\bf 182} (1988) 280.
\bibitem{Affleck} I. Affleck: Phys. Rev. Lett., {\bf 55} (1985) 1355.
\bibitem{Giamarchi-Schulz} T. Giamarchi and H. J. Schulz: 
	Phys. Rev. B {\bf 39} (1989) 4620.
\bibitem{Cardy86} J. L. Cardy: Nucl. Phys. B {\bf 270} [FS16] (1986) 186.
\bibitem{Reinicke} P. Reinicke: J. Phys. A {\bf 20} (1987) 5325.
\bibitem{Haldane} F. D. M. Haldane: Phys. Lett. {\bf 93A} (1983) 464;
	Phys. Rev. Lett. {\bf 50} (1983) 1153.
\bibitem{Botet} R. Botet and Jullien: Phys. Rev. B {\bf 27} 
	(1983) 613.
\bibitem{Sakai} T. Sakai and M. Takahashi: J. Phys. Soc. Jpn {\bf 59} 
	(1990) 2688. 
\bibitem{Yajima} M. Yajima and M. Takahashi: J. Phys. Soc. Jpn 
	{\bf 63} (1994) 3634. 
\bibitem{Kitazawa-Nomura-Okamoto} A. Kitazawa, K. Nomura, and 
	K. Okamoto: Phys. Rev. Lett. {\bf 76} (1996) 4038.
\bibitem{Kitazawa-Nomura} A. Kitazawa and K. Nomura: cond-mat/9705179, cond-mat/9705263
\bibitem{Oshikawa} M. Oshikawa: J. Phys.: Condens. Matt. {\bf 4}, (1992), 7469.
\bibitem{Cardy84} J. L. Cardy: J. Phys. A {\bf 17} (1984) L385.
\bibitem{Affleck86} I. Affleck: Phys. Rev. Lett. {\bf 56} (1986) 746.
\bibitem{Hatano93}
  N. Hatano and M. Suzuki,
  J. Phys. Soc. Jpn {\bf 62}, 1346 (1993).
\bibitem{Deisz93}
  J. Deisz, M. Jarrell, and D. L. Cox, 
  Phys. Rev. B {\bf 48} 10227 (1993).
\bibitem{Sun95}
  G. Sun,
  Phys. Rev. B {\bf 51}, 8370 (1995).
\bibitem{Nishiyama95}
  Y. Nishiyama, K. Totsuka, N. Hatano, and M. Suzuki, 
  J. Phys. Soc. Jpn {\bf 64}, 414 (1995).
\bibitem{Schollwock}
  U. Schollw\"ock and Th. Jolic{\oe}ur, 
  Europhys. Lett. {\bf 30}, 493 (1995).  
\bibitem{Yamamoto95}
  S. Yamamoto,
  Phys. Rev. Lett. {\bf 75}, 3348 (1995).
\bibitem{Qin95}
  S. Qin, T. K. Ng, and Z. B. Su,
  Phys. Rev. B {\bf 52}, 12844 (1995).
\bibitem{Qin97}
  S. Qin, Y. L. Liu, and L. Yu, 
  Phys. Rev. B {\bf 55}, 2721 (1997).
\bibitem{Nightingale86}
  M. P. Nightingale and H. W. Bl\"ote, Phys. Rev. B {\bf 33}, 650 (1986).
\bibitem{White92}
  S. R. White, Phys. Rev. Lett. {\bf 69}, 2863 (1992); Phys. Rev. B {\bf 48}, 
10345 (1993). 
\bibitem{Oshikawa-Yamanaka-Miyashita}
  M. Oshikawa, M. Yamanaka, and S. Miyashita, preprint (cond-mat/9507098).
\bibitem{Nomura-Okamoto} K. Nomura and K. Okamoto: J. Phys. A, 
	{\bf 27} (1994) 5773.  
\bibitem{Ziman-Schulz} T. A. L. Ziman and H. J. Schulz: Phys. Rev. Lett.,
	{\bf 59} (1987) 140.
\bibitem{Les-Houches} P. Ginsparg in {\em ``Fields, Strings, 
	and Critical Phenomena 	(1990) (Les Houches XLIX)} 
	ed. E. Br\'ezin and J. Zinn-Justin 
	(Amsterdam: North Holland).
\bibitem{Sutherland-Shastry} B. Sutherland and B. S. Shastry: Phys. Rev. Lett.,
	{\bf 65} (1990) 1833.
\bibitem{Fukui-Kawakami} T. Fukui and N. Kawakami: 
	J. Phys. Soc. Jpn. {\bf 65} (1996) 2824.
\bibitem{Kolb} M. Kolb: Phys Rev. B, {\bf 31}, (1985), 7494.
\bibitem{Fath-Solyom} G. F\'ath and J. S\'olyom: Phys Rev. B, 
	{\bf 47} (1993) 872.
\end{thebibliography}
